\documentclass[%
 aip,
 amsmath,amssymb,
 reprint,%
]{revtex4-1}

\usepackage{graphicx}
\usepackage{dcolumn}
\usepackage{bm}

\usepackage[utf8]{inputenc}
\usepackage[T1]{fontenc}
\usepackage{mathptmx}
\usepackage{etoolbox}

\makeatletter
\def\@email#1#2{%
 \endgroup
 \patchcmd{\titleblock@produce}
  {\frontmatter@RRAPformat}
  {\frontmatter@RRAPformat{\produce@RRAP{*#1\href{mailto:#2}{#2}}}\frontmatter@RRAPformat}
  {}{}
}%
\makeatother

\begin{document}

\title{Shock wave in series connected Josephson transmission line: Theoretical foundations and effects of resistive elements}

\author{Eugene Kogan}
\email{Eugene.Kogan@biu.ac.il}
\affiliation{Jack and Pearl Resnick Institute, Department of Physics, Bar-Ilan University, Ramat-Gan 52900, Israel}

\begin{abstract}
We  analytically study shock wave in the Josephson transmission line (JTL) in the presence of ohmic dissipation.
When ohmic resistors  shunt the Josephson junctions (JJ) or
are introduced in series with the ground capacitors the shock is broadened.
When
ohmic resistors are  in series with the JJ,
the shock remains  sharp, same as it was in the absence of  dissipation.
In all the cases considered, ohmic resistors don't influence the shock propagation velocity.
We study an alternative to the shock wave - an expansion fan - in the framework of the simple wave approximation  for the dissipationless JTL and formulate the generalization of the  approximation for the JTL with ohmic dissipation.

\end{abstract}

\date{\today}

\maketitle

\section{Introduction}
\label{intro}

The concept that in a nonlinear wave propagation system
the various parts of the wave travel with different
velocities, and that wave fronts (or tails) can sharpen
into shock waves, is deeply imbedded in the classical
theory of fluid dynamics \cite{whitham}.
The methods developed in that field can be profitably used
to study signal propagation in nonlinear transmission lines
\cite{hirota,kuek,kuek2,stykel,yamasaki,french,fatoorehchi,rangel0,nouri,neto,nikoo,silva,wang,rangel,kyuregyan,akem,fairbanks}.
In the early studies of shock waves in  transmission lines, the
origin of the nonlinearity was due to nonlinear capacitance
in the circuit \cite{landauer,peng,freeman,rabinovich}.

Interesting and potentially important examples of nonlinear transmission lines are circuits containing Josephson junctions (JJs) \cite{josephson} -
Josephson transmission lines (JTLs) \cite{barone,pedersen,tinkham,kadin}.
The unique nonlinear properties of JTLs allow to construct
soliton propagators,
microwave oscillators, mixers, detectors,
parametric amplifiers,  analog amplifiers \cite{pedersen,kadin,tinkham}.

Transmission lines
formed by in series connected
JJs were
studied beginning from  1990s, though much less than transmission lines
formed by JJs  connected in parallel \cite{solitons}.
However, the former began to attract quite a lot of attention  recently \cite{yaakobi,brien,macklin,kochetov,zorin,basko,dixon,goldstein}, especially
in connection with possible JTL traveling wave parametric amplification
\cite{yurke,yamamoto,beltram,hatridge,abdo,mutus,eichler,white,miano,pekker}.

Although numerical
calculations have revealed the existence of shock waves in series connected JTL
\cite{chen,mohebbi,katayama}, the analytical solutions have
not been successfully derived so far.
In the present paper we  study the propagation of shock waves in the JTLs analytically, paying especial attention to the influence of ohmic resistance which inevitably exists in the system.

The rest of the paper is constructed as follows. In Section~\ref{jtl}
we study the JTL equations in the absence of any ohmic dissipation. Shocks in the JTL with ohmic resistors  are studied in Section
\ref{ohm}.
In  Section~\ref{burgers} we generalize the  so called  simple wave approximation \cite{rabinovich,vinogradova}, which, for a dissipationless JTL,  reduces the wave equation to a pair of decoupled equations for right and left propagating waves, to the case of JTL with ohmic dissipation. We then  study the shock formation in the framework of this approximation.
Application of the results obtained in the paper and opportunities for their generalization are very briefly discussed in Section~\ref{discussion}.
We conclude in Section~\ref{conclusions}. In Appendix~\ref{lagham} we rederive the JTL equations in the framework of Lagrange and Hamilton approaches. In Appendix~\ref{spread}
we write down equation for the travelling wave in the discrete JTL.
In Appendix~\ref{integral} we consider  weak shocks in the JTL with large shunting capacitance.
In Appendix~\ref{exact} we study the symmetry of the modified Burgers equation proposed in Section~\ref{burgers} . In Appendix~\ref{ser} we study discontinuities formation in the JTL with ohmic resistors in series with the JJ.

\section{Dissipationless JTL}
\label{jtl}

\subsection{The JTL equations}
\label{sign}

\begin{figure}[h]
\includegraphics[width=\columnwidth]{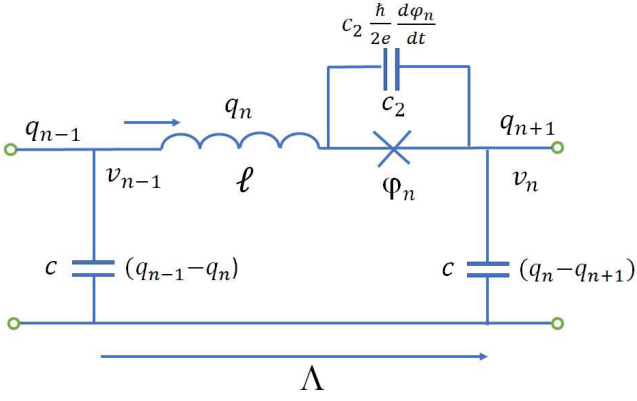}
\vskip -.5cm
\caption{Dissipationless  JTL. }
 \label{trans1}
\end{figure}

Consider the dissipationless model of JTL constructed from identical JJ, linear inductors and capacitors, and
indicated in Fig. \ref{trans1}.
We take
as dynamical variables  the phase differences  across the  JJ $\varphi_n$
and the charges which have passed through the  linear inductances $q_n$.
The  circuit equations are
\begin{subequations}
\label{ave7}
\begin{alignat}{4}
\frac{1}{c}\left(q_{n+1}-2q_{n}+q_{n-1}\right) &=  \ell\frac{d^2q_n }{d t^2}
+\frac{\hbar}{2e}\frac{d \varphi_n}{d t} \; ,\label{ave7a}\\
\frac{dq_n}{dt} &=   I_c\sin\varphi_n +c_2\frac{\hbar}{2e}\frac{d^2\varphi_n}{d t^2} \; ,\label{ave7b}
\end{alignat}
\end{subequations}
where   $\ell$ is the linear inductance, $c$ is the capacity between the wires, $c_2$ is the shunting capacity, and  $I_c$ is the critical current of the JJ.

Assuming smooth variance of  $\varphi_n$ and $q_n$ with $n$,
we can write down  (\ref{ave7}) in  the continuum approximation as
\begin{subequations}
\label{wave7x}
\begin{alignat}{4}
\frac{1}{C}\frac{\partial^2 q}{\partial x^2} &= L\frac{\partial^2q }{\partial t^2}+E_J\frac{\partial \varphi}{\partial t} \, ,\label{wave7xa}\\
\frac{\partial q}{\partial t} &= I_c\sin\varphi +C_2E_J\frac{\partial^2\varphi}{\partial t^2}\, ,\label{wave7xb}
\end{alignat}
\end{subequations}
where $\Lambda$  is the period of the line, $C= c/\Lambda$, $C_2=c_2\Lambda$,
$L=\ell/\Lambda$, and  $E_J=\hbar/(2e\Lambda)$.
In Appendix~\ref{lagham} we rederive Eqs. (\ref{ave7}) and (\ref{wave7x}) in the framework of Lagrange and Hamilton approaches.
Another  form of  (\ref{wave7xa}) can be obtained after we
 introduce  the voltage between the wires $v$:
\begin{subequations}
\label{wave7y}
\begin{alignat}{4}
\frac{\partial v}{\partial x} &= -L\frac{\partial^2q }{\partial t^2}-E_J\frac{\partial \varphi }{\partial t}  \, ,\label{wave7ya}\\
v &= -\frac{1}{C}\frac{\partial q}{\partial x}\, .\label{wave7yb}
\end{alignat}
\end{subequations}

\subsection{Waves and shock waves}
\label{wsw}

For $C_2=0$
equations (\ref{wave7y}) and (\ref{wave7xb}) take the form
\begin{subequations}
\label{wave79}
\begin{alignat}{4}
\frac{\partial  v}{\partial x} &= -L\frac{\partial }{\partial t}\left(I+
I_L \varphi\right) \, ,\label{wave79a}\\
C\frac{\partial v}{\partial t} &= -\frac{\partial I}{\partial x}\,,\label{wave79b}
\end{alignat}
\end{subequations}
where  $I_L=E_J/L$. In  (\ref{wave79a}) and (\ref{wave79b}) (and everywhere else in this paper) $I\equiv I_c\sin\varphi$.
If all variables in (\ref{wave79a}) and (\ref{wave79b})  are differentiable functions of $x$ and $t$, these equations can be combined in a single wave equation for $I$
\begin{eqnarray}
\label{wave77c}
\frac{\partial^2 I}{\partial x^2}-\frac{\partial}{\partial t}\left[\frac{1}{u^2(I)}\frac{\partial I}{\partial t}\right]=0\,,
\end{eqnarray}
where
\begin{eqnarray}
\label{ccc}
\frac{1}{u^2(I)}=CL_{eff}(I)\,,
\end{eqnarray}
and the effective inductance $L_{eff}(I)$ is
\begin{eqnarray}
L_{eff}=L\left(1+\frac{I_L}{I_c\cos\varphi}\right)=L\left(1+\frac{I_L}{(I_c^2-I^2)^{1/2}}\right)\;.
\end{eqnarray}
Note that $u(I)$ is the velocity propagation of  small amplitude disturbances on a homogeneous background, and that this velocity is given by  (\ref{ccc}) even for nonzero $C_2$, when the frequency of the disturbances
 $\omega\to 0$.

\begin{figure}[h]
\includegraphics[width=\columnwidth]{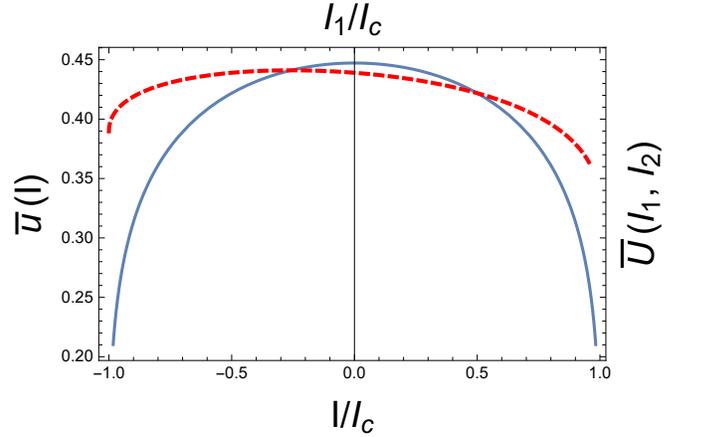}
\caption{Small amplitude disturbance  propagation velocity $\overline{u}(I)$
according to  (\ref{ccc}) (blue solid line) and shock propagation velocity $\overline{U}(I_1,I_2)$
according to  (\ref{velocity}) (red dashed line) for
$I_L=4I_c$ and  $I_2=.5I_c$. }
\label{velcity}
\end{figure}

In addition,  (\ref{wave79})
admits moving discontinuities  in the form
\begin{subequations}
\label{heavy}
\begin{alignat}{4}
I(x,t)&=I_1(x,t)H(-S(x,t))+I_2(x,t)H(S(x,t))\,,\\
v(x,t)&=v_1(x,t)H(-S(x,t))+v_2(x,t)H(S(x,t))\,,
\end{alignat}
\end{subequations}
where $H(x)$ is the Heaviside step function and $I_1,I_2$ ($v_1,v_2$) are functions with
continuous derivatives.
On substitution (\ref{heavy}) into (\ref{wave79})  we obtain (keeping only the singular terms)
\begin{subequations}
\label{hre}
\begin{alignat}{4}
\left[\frac{\partial S}{\partial x}\Delta v+
\frac{\partial S}{\partial t}L\left(\Delta I+I_L\Delta\varphi\right)\right]\delta(S(x,t))&=0\;,\\
\left(C\frac{\partial S}{\partial t}\Delta v+
\frac{\partial S}{\partial x}\Delta I\right)\delta(S(x,t))&=0
\end{alignat}
\end{subequations}
(everywhere in this paper $\Delta F\equiv F_2-F_1$, for any function $F$).
The velocity of the discontinuity propagation is
\begin{eqnarray}
U=-\left.\frac{\partial S/\partial t}{\partial S/\partial x}\right|_{S=0}\,
\end{eqnarray}
(in the simplest case $S(x,t)=t-x/U$).
Equation (\ref{hre}) therefore becomes
\begin{subequations}
\label{hren}
\begin{alignat}{4}
\Delta v-UL\left(\Delta I+I_L\Delta \varphi\right)&=0 \ ,\label{hrena}\\
UC\Delta v-\Delta I&=0 \ .\label{hrenb}
\end{alignat}
\end{subequations}
Eliminating $\Delta  v$  we
find
\begin{eqnarray}
\label{velocity}
\frac{1}{\overline{U}^2(I_1,I_2)}=1+I_L\frac{\Delta\varphi}{\Delta I}
=1+\frac{I_L\Delta\varphi}{I_c\Delta \sin\varphi}\,,
\end{eqnarray}
where $\overline{U}=U/u_T$, and  $u_T^2=1/LC$. The difference between $\overline{U}$ and $\overline{u}=u(I)/u_T$ is illustrated on Fig. \ref{velcity}.

In the particular case $L=0$, (\ref{velocity}) should be presented as
\begin{eqnarray}
\label{locity}
\frac{1}{U^2}=\frac{E_JC\Delta\varphi}{I_c\Delta \sin\varphi}\,.
\end{eqnarray}
In the symmetric case
\begin{eqnarray}
I_2=I_0=I_c\sin\frac{\Delta\varphi}{2},\hskip .5 cm I_1=-I_0=-I_c\sin\frac{\Delta\varphi}{2}\,,
\end{eqnarray}
 (\ref{locity}) takes the form
\begin{eqnarray}
\label{ocity}
\frac{1}{U^2}=\frac{E_JC}{I_0}\sin^{-1}\left(\frac{I_0}{I_c}\right)\,.
\end{eqnarray}
Expanding $\sin^{-1}$ in power series we obtain
\begin{eqnarray}
\label{ity}
\frac{1}{U^2}=\frac{E_JC}{I_c}\left(1+\frac{I_0^2}{6I_C^2}+\dots\right)\,.
\end{eqnarray}
The velocity, calculated in Ref. \cite{katayama} up to the second order in $I_0/I_c$, coincides with that given by (\ref{ity}) (also truncated to same order).

\subsubsection{Shock wave in the discrete JTL}

Actually, discontinuous solutions (\ref{heavy}) mean that the continuum approximation  is no longer adequate, and discrete model of the JTL, which resolves the discontinuity,
 should be considered. Let us start from rederiving  (\ref{velocity}) in the framework of the discrete model.

Assuming $c_2=0$, we can rewrite  (\ref{ave7a}) as
\begin{eqnarray}
\label{ave9}
\frac{1}{c}\left(q_{n+1}-2q_{n}+q_{n-1}\right)=\frac{d }{d t}\left(\ell I_c\sin\varphi_n
+\frac{\hbar}{2e}\varphi_n\right)  \,.
 \end{eqnarray}
Now let us sum up  (\ref{ave9}) between  the two points, one just ahead of the steep portion of the wave
front, and the other -  just behind it. The r.h.s. of the equation has
a time derivative only because of the motion of the steep
portion, and the slower changes due to the motion of the
parts of the wave with moderate slope can be neglected. Hence after the summation we obtain \begin{eqnarray}
\label{ave99}
\frac{\Lambda I_c}{cU}\Delta\sin\varphi=\frac{U}{\Lambda}\left(\ell I_c\Delta\sin\varphi
+\frac{\hbar}{2e}\Delta\varphi\right)\,,
 \end{eqnarray}
which is just  (\ref{velocity}).
(While calculating the l.h.s. of  (\ref{ave99}) we took into account  (\ref{ave7b}).)

Let us continue studying  (\ref{ave9}).
Introducing the new variable $i_n=dq_n/dt$ and  considering small amplitude disturbance of a uniform state
\begin{eqnarray}
\label{rel}
i_n=I_2-\Delta I\cdot x_n\;,
\end{eqnarray}
we can
linearise the problem with respect to $\Delta I$.
Introducing dimensionless time $\tau=2t/\sqrt{c\ell_{eff}}$,
where
\begin{eqnarray}
\ell_{eff}=\ell+\frac{\hbar/2e}{\sqrt{I_c^2-I_2^2}}\;,
\end{eqnarray}
we obtain from  (\ref{ave9})
\begin{eqnarray}
\label{1}
\frac{d^2x_n}{d\tau^2}=\frac{1}{4}(x_{n+1}-2x_n+x_{n-1})\, .
\end{eqnarray}

 We will consider  a signalling problem for a semi-infinite line $n\geq 0$.  The  problem is characterised by
the boundary  condition  $x_0(\tau)=1$ and   the initial conditions
$x_n(0)=\dot{x}_n(0)=0$ for $n\geq 1$.

\begin{figure}[h]
\includegraphics[width=\columnwidth]{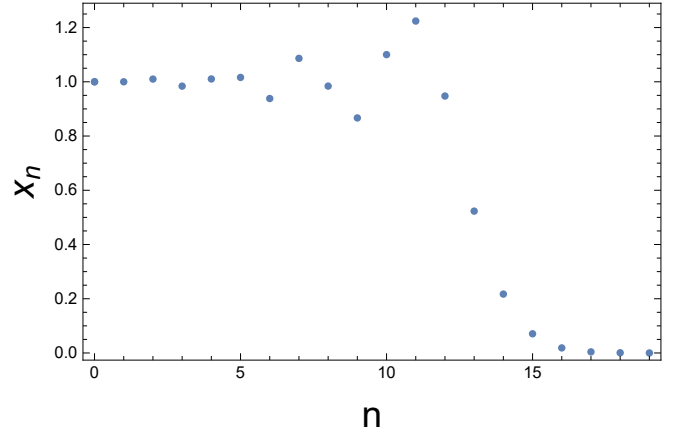}
\caption{A snapshot of the shock wave  given by  (\ref{b}) ($\tau=25$).}
\label{har}
\end{figure}

To solve  (\ref{1}) we will use the Laplace transform,  in the beginning following Ref. \cite{bulla}.
For a given time-dependent function $f(\tau)$, we define the
Laplace transform $F(s)={\cal L}\{f(\tau)\}$ as
\begin{equation}
\label{la}
 F(s) =
  \int_0^\infty {\rm d} \tau\, e^{-s\tau} f(\tau) \ .
\end{equation}
Laplace transforming  (\ref{1}) and  using
 the corollary of (\ref{la})
\begin{eqnarray}
f'(\tau)\Longleftrightarrow sF(s)-f(0)\ ,
\end{eqnarray}
we obtain the difference equation for $X_n(s)= {\cal L}\{x_n(t)\}$
\begin{equation}
\label{dif}
X_{n+1}-2(1+2s^2)X_n+X_{n-1}=0 \ ,
\end{equation}
with the boundary conditions $X_0(s)=1/s$, $\lim_{n\to\infty}X_n(s)=0$.
Solving (\ref{dif}):
\begin{equation}
\label{bro}
X_{n}(s)= \frac{1}{s}\left[1-2s\left(\sqrt{s^2+1}-s\right)\right]^n \; ,
\end{equation}
and taking into account  the known result \cite{abram}
\begin{eqnarray}
(\sqrt{s^2+1}-s)^k\Longleftrightarrow \frac{k}{\tau}J_k(\tau)\;,
\end{eqnarray}
where $J_k$ is the Bessel function,
we obtain
\begin{eqnarray}
\label{b}
x_n(\tau)=1+\sum_{k=1}^nC^n_k(-1)^k2^kk\frac{d^{k-1}}{d\tau^{k-1}}
\left(\frac{J_k(\tau)}{\tau}\right).
\end{eqnarray}
Using the recurrence relation
\begin{eqnarray}
2\frac{d}{d\tau}J_k(\tau)=J_{k-1}(\tau)-J_{k+1}(\tau)\; ,
\end{eqnarray}
we  can present any $x_n$ as a linear combination of Bessel functions. A snapshot of the shock wave  given by  (\ref{b}) is presented on Fig. \ref{har}.

The input voltage is
\begin{eqnarray}
v_0(t)&=&\frac{1}{c}\left[q_0(t)-q_1(t)\right]=\frac{\Delta I}{c}\int_0^tdt[x_1(t)-1]\nonumber\\
&=&\sqrt{\frac{\ell_{eff}}{c}}\Delta I\int_0^{\tau}d\tau\frac{J_1(\tau)}{\tau}\ .
\end{eqnarray}
In particular, using known formula for the integral from Bessel function \cite{prudnikov},
we obtain  an expected result
\begin{eqnarray}
v_0(\infty)=\sqrt{\frac{\ell_{eff}}{c}}\Delta I=\sqrt{\frac{L_{eff}}{C}}\Delta I=Z_{eff}\Delta I\;.
\end{eqnarray}

These results  show how the (quasi) discontinuous shock wave can be generated. If there is  a semi-infinite JTL with  the constant current source at the end (and  in  a stationary state, with the input voltage being equal to zero),  and  then suddenly  the current of the  source changes to another constant value, the shock wave will start to propagate.

In order to compute the inverse Laplace transform one
can either  use
 correspondence tables, like we did above, or compute the Bromwich integral.
More specifically, there exists the following theorem.

If the function $F(s)$ is analytic in the half plane Re $s>s_0$, goes to zero when $|s|\to \infty$ in any half plane Re $s\geq a>s_0$ uniformly with respect to arg $s$ and the integral
\begin{eqnarray*}
\int_{a-i\infty}^{a+i\infty}F(s)ds
\end{eqnarray*}
absolutely converges, then $F(s)$ is the Laplace image of the function
\begin{eqnarray}
\label{brom}
f(\tau)=\frac{1}{2\pi i}\int_{a-i\infty}^{a+i\infty}ds\,e^{s\tau}F(s)
\end{eqnarray}
(the integration is done along the vertical line Re$(s) = a$ in the complex plane).

\begin{figure}[h]
\includegraphics[width=\columnwidth]{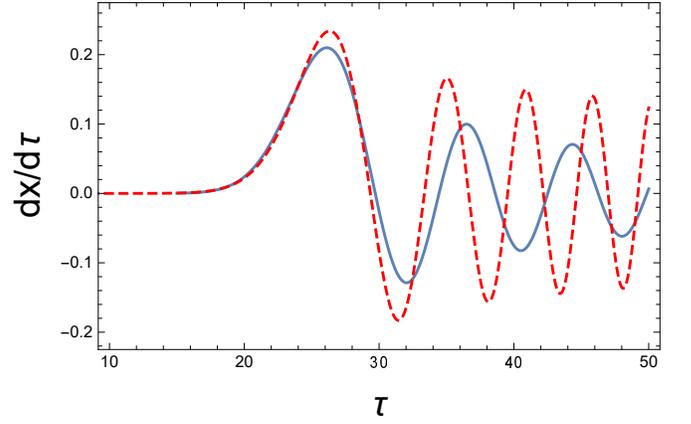}
\caption{Equation (\ref{b2}) (solid blue line) and  (\ref{bro4}) (dashed red line) for $n=12$.}
\label{comp}
\end{figure}

From  (\ref{b}) follows
\begin{eqnarray}
\label{b2}
\frac{dx_n}{d\tau}=\sum_{k=1}^nC^n_k(-1)^k2^kk\frac{d^{k}}{d\tau^{k}}
\left(\frac{J_k(\tau)}{\tau}\right)\,.
\end{eqnarray}
We will try to get more explicit analytic result for the quantity using Bromwich integral.
From  (\ref{bro}) we obtain
\begin{eqnarray}
\label{bro2}
&&\frac{dx_{n}}{d\tau}= \frac{1}{2\pi i}\int_{a-i\infty}^{a+i\infty}ds \nonumber\\
&&\exp\left\{s\tau+n\ln\left[1-2s\left(\sqrt{s^2+1}-s\right)
\right]\right\}\; .
\end{eqnarray}
Expanding the logarithm in  (\ref{bro2}) with respect to $s$ and keeping only the lowest order terms we get
\begin{eqnarray}
\label{bro3}
\frac{dx_{n}}{d\tau}= \frac{1}{2\pi i}\int_{a-i\infty}^{a+i\infty}ds
\exp\left[s(\tau-2n)+\frac{n}{3}s^3\right]\; .
\end{eqnarray}
The contour of integration in  (\ref{bro3}) can be deformed so that it will start  at the point at infinity with argument $-\pi/3$ and will end at the point at infinity with argument $\pi/3$. Hence the integral  determines the Airy function \cite{abram}
\begin{eqnarray}
\label{bro4}
\frac{dx_{n}}{d\tau}=n^{-1/3} \text{Ai}\left[n^{-1/3}(2n-\tau)\right]\; .
\end{eqnarray}
Equation (\ref{bro4}) describes the shock front at $n\sim \tau/2$,
exponential decrease of the signal with increasing $n$ for $n>\tau/2$, and oscillations and power law decrease of the signal with decreasing $n$ for $n<\tau/2$.
Equations (\ref{b2})  and  (\ref{bro4}) are plotted on Fig. \ref{comp}. Comparison of the graphs shows that the approximate formula (\ref{bro4}) correctly describes front of the shock wave, but not its wake.

In the linear approximation to  (\ref{ave9}), as we see from  (\ref{bro4}),
the shock wave   spreads indefinitely (because of the dispersion present in the system).
An attempt to understand
whether the nonlinear terms in the equation  will stop the spreading, is presented in Appendix~\ref{spread}.

\section{Shock waves in the JTL with ohmic dissipation}
\label{ohm}

\begin{figure}[h]
\includegraphics[width=\columnwidth]{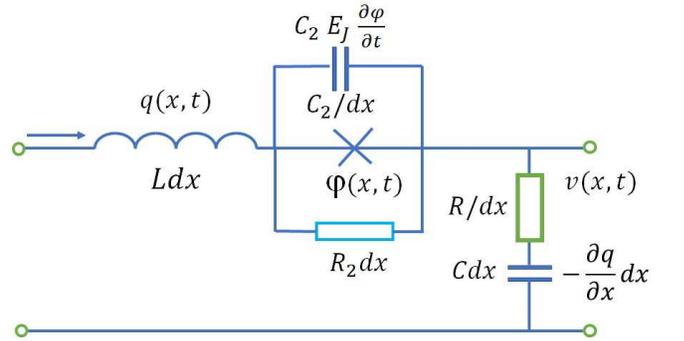}
\caption{JTL with ohmic resistors
shunting the JJ and  in series with the ground capacitor. }
 \label{trans3}
\end{figure}

Let us add to the JTL ohmic resistors, thus  considering  transmission line
presented on Fig. \ref{trans3}.
In this case  (\ref{wave7xb}) changes to
\begin{eqnarray}
\label{wave7ynb}
\frac{\partial q}{\partial t} = I_c\sin\varphi+\frac{E_J}{R_2}\frac{\partial \varphi}{\partial t}+C_2E_J\frac{\partial^2 \varphi}{\partial t^2}\,,
\end{eqnarray}
and  (\ref{wave7yb}) changes to
\begin{eqnarray}
\label{wave7yn}
v = -\frac{1}{C}\frac{\partial q}{\partial x}-R\frac{\partial^2 q}{\partial t\partial x}\,.
\end{eqnarray}

\subsection{The travelling waves}
\label{gc}

The system of equations  (\ref{wave7ya}), (\ref{wave7ynb}), (\ref{wave7yn})  has a set of particularly simple solutions,
called traveling waves, when  dependence of all the quantities upon $x,t$ is the dependence upon a single parameter $t-x/U$ (for the sake of definiteness, we consider  right going waves).
These solutions satisfy ordinary differential equation, which can be obtained after eliminating $v$ and $q$:
\begin{eqnarray}
\label{avec2}
RCC_2L \frac{d^3 \varphi}{d t^3}+\left[\frac{RC}{R_2}
+C_2\left(1-\overline{U}^2\right)\right]L \frac{d^2 \varphi}{d t^2} \nonumber\\
+\left[\frac{L}{R_2}\left(1-\overline{U}^2\right)+\frac{RCI_c}{I_L}\cos\varphi\right]\frac{d \varphi}{dt}\nonumber\\
+\frac{I_c}{I_L}\sin\varphi\left(1-\overline{U}^2\right)
-\overline{U}^2\varphi+A=0\,,
\end{eqnarray}
where  $A$ is an arbitrary constant (we integrated once).

We are looking for a solution which tends to constants at infinity
\begin{eqnarray}
\label{boundary}
\lim_{t\to -\infty}I= I_2,\;\;\lim_{t\to +\infty}I=I_1\,.
\end{eqnarray}
For given $I_1$ and   $I_2$, the  parameters $U,A$ must satisfy
\begin{eqnarray}
\label{uu}
I_i-\overline{U}^2\left(I_i+I_L\varphi_i\right)+A=0\,, \;\;\;i=1,2\,.
\end{eqnarray}
Eliminating $A$ we recover  (\ref{velocity}).
We see that the relation between the shock velocity and the values of current on both sides of the shock is dissipation independent.

Equation (\ref{avec2}) with  the boundary conditions (\ref{boundary}) can be presented as
\begin{eqnarray}
\label{bve0}
RCC_2L\frac{d^3 \varphi}{d t^3}+\left[\frac{RCL}{R_2}
+C_2L\left(1-\overline{U}^2\right)\right]\frac{d^2 \varphi}{d t^2}\nonumber\\
+\left[\frac{L}{R_2}\left(1-\overline{U}^2\right)+\frac{RCI_c}{I_L}\cos\varphi\right]\frac{d \varphi}{dt}=M(\varphi)\,,
\end{eqnarray}
where
\begin{eqnarray}
\label{bv}
M(\varphi)&=&\left[\varphi\cdot\Delta\sin\varphi-\sin\varphi\cdot\Delta\varphi +\varphi_2\sin\varphi_1-\varphi_1\sin\varphi_2\right]\nonumber\\
&\cdot&\frac{I_c}{I_c\Delta\sin\varphi+I_L\Delta\varphi}\,.
\end{eqnarray}
Note that $M(\varphi_1)=M(\varphi_2)=0$, and that for weak shock  ($|I_2-I_1| \ll |I_2|$),
$M(\varphi)$ is simplified to
\begin{eqnarray}
\label{bve76}
M(\varphi)=\overline{u}_0^2\frac{\tan\varphi_0}{2}(\varphi-\varphi_1)(\varphi-\varphi_2)\,.
\end{eqnarray}

Equation (\ref{bve0})  is very close to that describing dc driven resistively shunted self-resonant Josephson tunnel junction \cite{imry,kruger,likharev}.

\subsection{Newtonian analogy}
\label{newton}

\subsubsection{Stability of the equilibrium points}

\begin{figure}[h]
\includegraphics[width=.6\columnwidth]{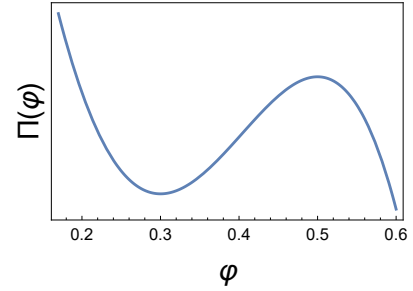}
\caption{Potential energy of the fictitious particle  (arbitrary units); $\varphi_1=.3,\varphi_2=.5$.}
\label{well}
\end{figure}

If  $\sqrt[3]{RCC_2L}$ is much less than all the other time scales in the problem, the term with the third derivative  in  (\ref{bve0}) can be discarded, and  the latter takes the form
\begin{eqnarray}
\label{dvec}
\frac{d^2 \varphi}{d\tau^2}+(\gamma_2+\gamma_1\cos\varphi)
\frac{d \varphi}{d\tau}+\frac{d\Pi(\varphi)}{d\varphi}=0\,,
\end{eqnarray}
where
\begin{eqnarray}
\label{change}
\tau=\frac{t}{T},\;\;\;
\gamma_2&=&\frac{L}{R_2T}\left(1-\overline{U}^2\right),\;\;\;\gamma_1=\frac{RCI_c}{I_LT},\nonumber\\
T^2&=&\frac{RCL}{R_2}+C_2L\left(1-\overline{U}^2\right)\,,
\end{eqnarray}
and
\begin{eqnarray}
\Pi(\varphi)&=&\Big[-\frac{\varphi^2}{2}\cdot\Delta\sin\varphi+(1-\cos\varphi)\cdot\Delta\varphi
\\
&-&\varphi\cdot(\varphi_2\sin\varphi_1-\varphi_1\sin\varphi_2)\Big]
\cdot\frac{I_c}{I_c\Delta\sin\varphi+I_L\Delta\varphi}\,.\nonumber
\end{eqnarray}
Equation (\ref{dvec}) describes motion of a Newtonian particle
 in the potential well
$\Pi(\varphi)$, shown on Fig. \ref{well}, and
in the presence of friction force (the term proportional to
$d \varphi/d\tau$).
The particle (asymptotically) starts in the upper equilibrium position and finishes in the lower equilibrium position.

The values $\varphi_1$ and $\varphi_2$ enter  into  (\ref{velocity}) in a symmetrical way.  However, due to ohmic resistance,
inevitably present in the system, only one direction of shock propagation is possible.
The potential energy of the fictitious particle corresponding to the state before the shock should be higher than that behind the shock. This condition can be enhanced even further.
The potential energy shouldn't have any local extrema between $\varphi_1$ and $\varphi_2$ (local extremum  means splitting of a single shock into two.)
Thus  the potential energy   should have at $\varphi_2$ local maximum, and at $\varphi_1$ - local minimum. This is equivalent to inequalities
\begin{eqnarray}
\label{inequality3}
u(I_1)>U(I_1,I_2)>u(I_2)\,,
\end{eqnarray}
which reflect the  well known  fact: the shock velocity is lower than the sound velocity in the region behind the shock, but higher than the sound velocity   in the region before the shock \cite{whitham}. Actually, the stability analysis of the equilibrium points can be
performed for  (\ref{bve0}) with the same result.

Because $\sin\varphi$ is concave downward for
$0<\varphi<\pi/2$, and concave upward for $-\pi/2<\varphi<0$, $M(\varphi)$ can not have zeros between $\varphi_1$ and $\varphi_2$ having the same sign, thus there can exist shock
between any pair of currents of the same sign.
On the other hand, the inequalities (\ref{inequality3}) pose limitations on the values of positive and negative currents, between which a single shock can exist.
So returning to Fig. \ref{velcity} we understand, that  the red dashed curve  inside the dome describes shocks, for which $I_2$ is the current before the shock. The red dashed curve to the right of the dome and the red dashed curve to the left of the dome which lies below $\overline{u}(I_2)$
describe the shocks, for which $I_2$ is the current after the shock.

\subsubsection{The shock profile}

Though  (\ref{dvec}) is non integrable analytically in the general case, qualitatively the nature of the motion from one equilibrium position to the other is clear (at least for $\gamma\gg 1$ and $\gamma\ll 1$).
In the former regime the particle moves monotonically from one equilibrium position to the other,
in the latter - the particle
oscillates in the potential well, and  weak friction  leads to slow decrease of the oscillations amplitude with time.

\begin{figure}[h]
\includegraphics[width=\columnwidth]{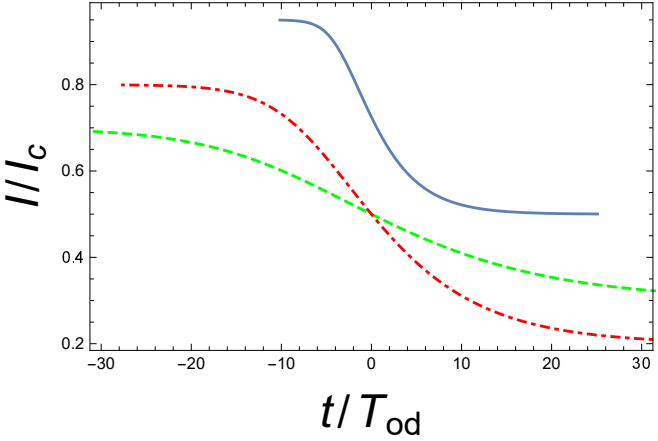}
\caption{Shock profile according to  (\ref{combb}) for  $I_L=.5I_c$.
Blue solid line corresponds to $I_1=.5I_c$, $I_2=.95I_c$, red dot-dashed line - to
$I_1=.2I_c$, $I_2=.8I_c$, green dashed line - to
$I_1=.3I_c$, $I_2=.3I_c$, $T_{od}=RCI_c/I_L$.}
\label{shock}
\end{figure}

 If $C_2=0$ and $\sqrt{L/(RR_2C)}\ll 1$, the terms with $\gamma_2$ and the second derivative  in the l.h.s. of (\ref{dvec}) can be discarded.
If $C_2=0$ and $\sqrt{L/(RR_2C)}\gg 1$, the terms with $\gamma_1$ and the second derivative  can be discarded.  The resulting equations describe motion of the strongly overdamped particle and can be easily integrated analytically.
 We obtain
\begin{subequations}
\label{comb}
\begin{alignat}{4}
t&=\frac{RCI_c}{I_L}\int\frac{\cos\varphi d \varphi}{M(\varphi)}, \hskip 1.1cm \sqrt{\frac{L}{RR_2C}}\ll 1\,, \label{combb}\\
t&=\frac{L}{R_2}\left(1-\overline{U}^2\right)\int\frac{d \varphi}{M(\varphi)}, \hskip .5cm \sqrt{\frac{L}{RR_2C}}\gg 1 \, .\label{comba}
\end{alignat}
\end{subequations}
In both cases  the shape of the shock depends only upon $\varphi_1$ and $\varphi_2$ and is independent upon the parameters of the transition line.
Equation (\ref{combb})  is presented graphically on Fig. \ref{shock}.
For weak shock  in both cases we obtain
\begin{eqnarray}
\label{cot}
I=I_0-\frac{\Delta I}{2}\tanh(\alpha t)\,,
\end{eqnarray}
where $I_0=(I_1+I_1)/2$, and
\begin{subequations}
\label{omb}
\begin{alignat}{4}
\alpha&=\frac{\Delta u}{2u_0}\frac{1}{RC}, \hskip 3cm \sqrt{\frac{L}{RR_2C}}\ll 1 \label{ombb}\\
\alpha&=\frac{I_L}{I_c\cos\varphi_0}\frac{\Delta u}{2u_0}\frac{R_2}{L\left(1-\overline{U}^2\right)}, \hskip .5cm \sqrt{\frac{L}{RR_2C}}\gg 1 \, .\label{omba}
\end{alignat}
\end{subequations}

\begin{figure}[h]
\includegraphics[width=\columnwidth]{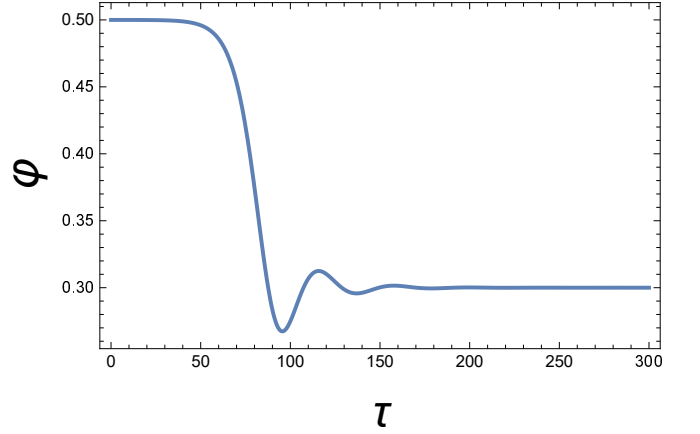}
\caption{Shock profile according to  (\ref{dvec}); $I_L=.5I_c,\varphi_1=.3,\varphi_2=.5,\gamma_2=.1$, $\gamma_1=0$.}
\label{SShock}
\end{figure}

\begin{figure}[h]
\includegraphics[width=\columnwidth]{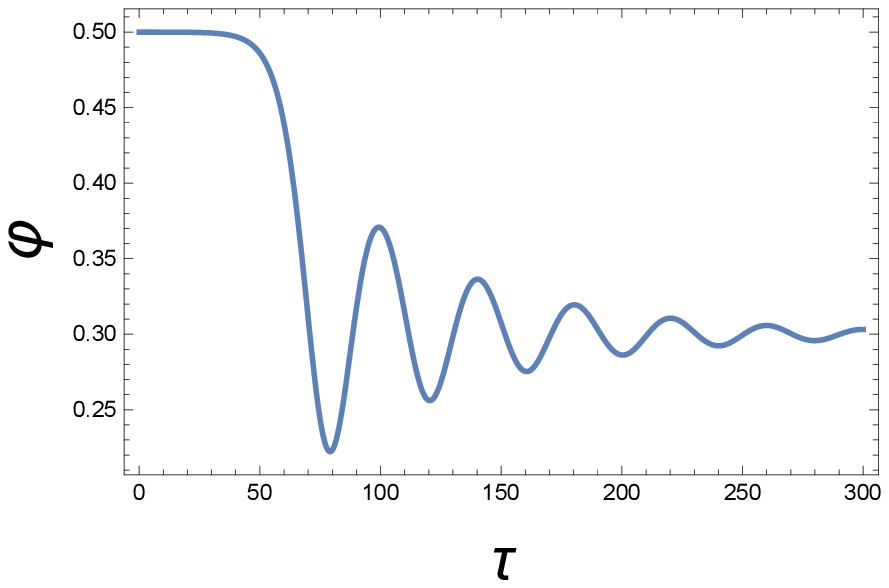}
\caption{Shock profile according to  (\ref{dvec}); $I_L=.5I_c,\varphi_1=.3,\varphi_2=.5,\gamma_2=.03$, $\gamma_1=0$.}
\label{Shock}
\end{figure}

Consider now the case of $R=0$,  $C_2\gg L/R_2^2$ (which corresponds to $\gamma_2\ll 1$).
The results of integration of  (\ref{dvec}) in this case are presented on Figs.  \ref{SShock} and \ref{Shock}.
  The phase (current) in the shock wave oscillates, in strong contrast to monotonous change in the case of zero shunting capacitance, presented on Fig. \ref{shock}.

\subsection{Weak damping: the method of time averaging}
\label{je}

When the term with the third derivative in  (\ref{bve0}) is kept, the latter can be written  as
\begin{eqnarray}
\label{dvec7}
\beta\frac{d^3 \varphi}{d\tau^3}+\frac{d^2 \varphi}{d\tau^2}+\gamma(\varphi)\frac{d \varphi}{d\tau}+\frac{d\Pi(\varphi)}{d\varphi}=0\,,
\end{eqnarray}
where $\beta=RCC_2L/T^3$. Equation (\ref{dvec7})  describes jerky  \cite{linz,sprott} particle. Analysis of the local stability of the fixed point $\varphi=\varphi_1$ is exactly the same as it was in Section \ref{newton}. Global stability of the point is less obvious. In any case, in this Section
we will consider   the regime $C_2\gg L/R_2^2,R^2C^2/L$  (which corresponds to $\beta,\gamma\ll 1$), where the situation with the global stability is clear (see (\ref{w3b}) and the sentence immediately after it).

We will use  the method of time averaging, which we formulate below.
We assume that we know the undamped solution
 $\varphi_{ud}(\tau;{\cal E})$,  satisfying equation
\begin{eqnarray}
\label{con}
\frac{1}{2}\left(\frac{d\varphi_{ud}}{d\tau}\right)^2+\Pi(\varphi_{ud})={\cal E}\,,
\end{eqnarray}
and express    damped oscillations as
\begin{eqnarray}
\label{time}
\varphi(\tau)=\varphi_{ud}(\tau;{\cal E}(\tau))\,.
\end{eqnarray}
To find ${\cal E}(\tau)$, notice that
from  (\ref{dvec}) follows
\begin{eqnarray}
\label{w3}
\frac{d}{d\tau}\left[\frac{1}{2}\left(\frac{d\varphi}{d\tau}\right)^2+\Pi(\varphi)\right]
=-\gamma(\varphi)\left(\frac{d \varphi}{d\tau}\right)^2-\beta\frac{d \varphi}{d\tau}\frac{d^3 \varphi}{d\tau^3}\,.\nonumber\\
\end{eqnarray}
Ignoring terms of the order of $\beta \gamma$ and $\beta^2$,  (\ref{w3}) can be written as
\begin{eqnarray}
\label{w3b}
\frac{d}{d\tau}\left[\frac{1}{2}\left(\frac{d\varphi}{d\tau}\right)^2+\Pi(\varphi)\right]
=-\Gamma(\varphi)\left(\frac{d \varphi}{d\tau}\right)^2\,,\nonumber\\
\end{eqnarray}
where $\Gamma(\varphi)=\gamma(\varphi)+\beta dM(\varphi)/d\varphi$.
Note, that  $\Gamma(\varphi)$ is positive for any $\varphi$. For example, when $R=\infty$,
\begin{eqnarray}
\Gamma(\varphi)=\frac{RCC_2LI_c}{T^3I_L}\left(1-\overline{U}^2\right)
\frac{(I_c\cos\varphi+I_L)\Delta\sin\varphi}{I_c\Delta\sin\varphi+I_L\Delta\varphi}\,.
\end{eqnarray}
We'll  assume that  ${\cal E}(\tau)$ satisfies equation
\begin{eqnarray}
\label{eee}
\frac{d {\cal E}}{d\tau}=-\left<\Gamma(\varphi)\left(\frac{d \varphi}{d\tau}\right)^2\right>_{ud}
\equiv -{\cal A}({\cal E})\,,
\end{eqnarray}
where the averaging is with respect to the period of the undamped oscillation with the energy ${\cal E}$.
The   averaging in (\ref{eee}) is performed as
\begin{eqnarray}
\label{eeey}
\frac{1}{T}\int_{0}^{T}\Gamma(\varphi)\left(\frac{d \varphi_{ud}}{dt}\right)^2dt
=\frac{2\int d\varphi \Gamma(\varphi)\sqrt{{\cal E}-\Pi(\varphi)}}
{\int d\varphi/\sqrt{{\cal E}-\Pi(\varphi)}}\,,
\end{eqnarray}
the limits of integration in both integrals being found from the equation
${\cal E}-\Pi(\varphi)=0$.
Integrals defining ${\cal A}({\cal E})$ being calculated, the
solution of (\ref{eee})
\begin{eqnarray}
\label{ee}
\tau=-\int\frac{d {\cal E}}{{\cal A}({\cal E})}\,,
\end{eqnarray}
together with   (\ref{time}), gives parametric representation of the particle motion. In Appendix \ref{integral} we'll see how  all this works for weak shocks.

\subsection{The JTL with ohmic resistor
in series with the JJ}
\label{series}

\begin{figure}[h]
\includegraphics[width=\columnwidth]{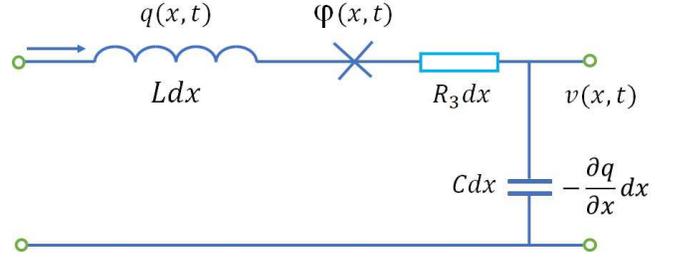}
\caption{JTL with ohmic resistor
in series with the JJ. }
 \label{trans5}
\end{figure}

Let us now introduce ohmic resistor in a way different
from that considered   previously,
constructing the transmission line presented on Fig.  \ref{trans5}.
In this case  (\ref{wave79}) changes to
\begin{subequations}
\label{ave79}
\begin{alignat}{4}
\frac{\partial  v}{\partial x} &= -L\frac{\partial }{\partial t}\left(I+
I_L \varphi\right)-R_3I \, ,\\
C\frac{\partial v}{\partial t} &= -\frac{\partial I}{\partial x}
\end{alignat}
\end{subequations}

The analysis of moving discontinuities presented in Section \ref{wsw} can be repeated verbatim,
hence (i) the discontinuous solutions are allowed for  (\ref{ave79}), (ii)
 the velocity of their propagation  is given by  (\ref{velocity}).
Ohmic resistors in the present case neither lead to finite width of shocks, nor influence the velocity of their propagation.
Notice, that in the presence of the higher order derivatives in the JTL equations,
which is the case when ohmic dissipation is taken into account,
the terms with
$\delta'(x-Ut)$ will appear if we substitute discontinuous functions of coordinates, and there are no other  that singular term to balance it.

\section{The simple wave approximation}
\label{burgers}

\subsection{The dissipationless JTL}
\label{form}

Let us start from the dissipationless transmission line, described by  (\ref{wave77c}).
The simple wave approximation \cite{rabinovich,vinogradova} for the equation may be obtained by changing by brute force  (\ref{wave77c}) into two decoupled equations for right and left going waves
\begin{eqnarray}
\label{ve8}
\frac{\partial I}{\partial x}\pm \frac{1}{u(I)}\frac{\partial I}{\partial t}=0\,.
\end{eqnarray}
Equation (\ref{ve8}), in distinction to   (\ref{wave77c}),
can be easily solved analytically \cite{whitham,billingham}.

Consider the signalling problem for $x\geq 0$, characterised by the initial and boundary
conditions
\begin{subequations}
\label{bo}
\begin{alignat}{4}
I(x,0)&=I_2 \, ,\\
I(0,t)&=I_1 \, .
\end{alignat}
\end{subequations}
The solution of (\ref{ve8}) containing the shock is
\begin{eqnarray}
\label{s}
I(x,t)=\left\{\begin{array}{ll} I_2,\; & \text{for}\;\;\;x>U(I_1,I_2)t, \\
I_1,\; & \text{for}\;\;\;0<x<U(I_1,I_2)t.\end{array}\right.
\end{eqnarray}
However,  the shock can exist only provided $|I_1|<|I_2|$.
For $|I_1|>|I_2|$ the solution of  (\ref{ve8}) contains an expansion fan \cite{whitham,billingham}
\begin{eqnarray}
I(x,t)=\left\{\begin{array}{ll} I_2,\; & \text{for}\;\;\;x>u(I_2)t, \\
{\cal I}(x/u_Tt),\; & \text{for}\;\;\;u(I_1)t\leq x\leq u(I_2)t, \\
I_1,\; & \text{for}\;\;\;0<x<u(I_1)t,\end{array}\right.
\end{eqnarray}
where the function  ${\cal I}$ is obtained by  inverting (\ref{ccc})
\begin{eqnarray}
{\cal I}(\overline{u})
=\left[I_c^2-\frac{I_L^2\overline{u}^4}{(1-\overline{u}^2)^2}\right]^{1/2}\,.
\end{eqnarray}

\subsection{Shock formation in the JTL with ohmic dissipation}
\label{jd}

When ohmic resistor in series with the ground capacitor is additionally taken into account,
 (\ref{wave77c}) is modified to
\begin{eqnarray}
\label{wave00}
\frac{\partial^2 }{\partial x^2}\left(I+RC\frac{\partial I}{\partial t}\right)
=\frac{\partial }{\partial t}\left[\frac{1}{u^2(I)}\frac{\partial I}{\partial t}\right]\,.
\end{eqnarray}
Following the example of Section \ref{form},  we attempt to
 take the "square root" of the operator, acting upon $I$  in the l.h.s. of  (\ref{wave00}), and postulate  that right  and left going waves  satisfy decoupled  equations
\begin{eqnarray}
\label{ve9}
&&\frac{\partial I}{\partial t}\pm u(I)\frac{\partial }{\partial x}
\left(I+\nu\frac{\partial I}{\partial t}\right)=0\,,
\end{eqnarray}
where $\nu=RC/2$.
Equations (\ref{ve9}) are the generalization of the simple wave approximation to the case of JTL with ohmic dissipation.

For the traveling right going wave, from  (\ref{ve9}) we obtain
\begin{eqnarray}
\label{ve94}
\frac{1}{U}\frac{d }{dt}
\left(I+\nu\frac{d I}{d t}\right)-\frac{1}{ u(I)}\frac{d I}{d t}=0\,.
\end{eqnarray}
Integrating with respect to $t$ from $-\infty$ to $+\infty$ and taking into account  the boundary conditions (\ref{boundary}), we obtain
\begin{eqnarray}
\label{wrong}
\frac{1}{U}=\frac{1}{\Delta I}\int_{I_1}^{I_2}  \frac{dI}{u(I)}\,.
\end{eqnarray}
More explicitly, (\ref{wrong}) is
\begin{eqnarray}
\label{veloc}
\frac{1}{\overline{U}(I_1,I_2)}=\frac{1}{\Delta I}\int_{I_1}^{I_2} dI\left[1+\frac{I_L}{(I_c^2-I^2)^{1/2}}\right]^{1/2}\,.
\end{eqnarray}
Equation (\ref{veloc}) is slightly different from  the exact  (\ref{velocity}),
but the results are very close. We didn't plot the curve given by  (\ref{veloc}) on Fig. \ref{velcity}, because it would absolutely merge with the curve given by (\ref{velocity}).
Another argument, which convinces us in validity of  (\ref{ve9}), is the fact that
the weak shock profile (\ref{cot}), with $\alpha$ given  by   (\ref{ombb}), is a solution of (\ref{ve9}).

Equation (\ref{ve9}) hopefully is able to describe the shock formation for the signalling problem. To make this task easier, we propose to additionally simplify  (\ref{ve9}) to
\begin{eqnarray}
\label{proba22}
\frac{\partial u}{\partial t} \pm u\frac{\partial }{\partial x}\left(u +\nu \frac{\partial u}{\partial t}\right)=0\,.
\end{eqnarray}
Equation (\ref{proba22}) may be called the modified Burgers equation (mBE).
In Appendix~\ref{exact} we study the symmetry of the equation.

Now let us consider the JTL with ohmic resistor in series with the JJ.
From  (\ref{ave79}) follows
\begin{eqnarray}
\label{wave77cd}
\frac{\partial^2 I}{\partial x^2}-\frac{\partial}{\partial t}\left[\frac{1}{u^2(I)}\frac{\partial I}{\partial t}+R_3CI\right]=0.
\end{eqnarray}
The
generalization of the simple wave approximation  to this case is
\begin{eqnarray}
\label{ve9d}
\frac{\partial I}{\partial t} +\mu u^2(I)I\pm u(I)\frac{\partial I}{\partial x}
=0\,,
\end{eqnarray}
where $\mu=R_3C/2$.
Discontinuities formation for the solutions of  (\ref{ve9d}) is studied in Appendix~\ref{ser}.

\section{Discussion}
\label{discussion}

We hope that the results obtained in the paper are  applicable to kinetic inductance based traveling wave parametric amplifiers based on a coplanar waveguide architecture.
Onset of shock-waves in
such amplifiers is an undesirable phenomenon. Therefore, shock
waves in various JTL should be further studied, which was one of motivations of the present work.

Recently,  quantum mechanical description of JTL in general and parametric amplification in such lines in particular started to be developed, based on
quantisation techniques in terms
of discrete mode operators \cite{reep},  continuous  mode operators \cite{fasolo},
 a Hamiltonian approach  in the Heisenberg and interaction pictures \cite{greco},
 or the quantum Langevin method \cite{yuan}
It  would be interesting to understand in what way the results of the present paper are changed by quantum mechanics.
Particularly interesting looks studying  of
quantum ripples over a semi-classical shock \cite{glazman} and fate of quantum shock waves at late times \cite{glazman2}.

\section{Conclusions}
\label{conclusions}

We have analytically calculated the velocity of propagation and structure of shock waves in the transmission line constructed from the JJ, linear inductors,  capacitors and ohmic resistors. In the absence of  ohmic dissipation the shocks are  sharp. As such they remain when
ohmic resistors are introduced in series with the JJ and  linear inductors.
When ohmic resistors  shunt the JJ or  are in series with the ground capacitors, the shocks are broadened.
The shock width  is inversely proportional to the resistance shunting JJ, or proportional to the resistance in series with the ground capacitor.
 In all the cases considered, ohmic resistors (and shunting capacitors) don't influence the shock propagation velocity.
We formulate the simple wave approximation for the JTL with ohmic dissipation and
 study an alternative to the shock wave - an expansion fan - in the framework of this approximation.

\begin{acknowledgments}

Discussions with  R. Bulla, L. Friedland, L. Glazman, M. Goldstein,  H. Katayama, N. Kwidzinski, K. O'Brien, T. H. A. van der Reep, B. Ya. Shapiro, A. Sinner, F. Vasko, M. Yarmohammadi, R. Zarghami and A. B. Zorin are gratefully acknowledged.

\end{acknowledgments}

\section*{Data Availability Statement}

The data that support the findings of this study are available from the corresponding author
upon reasonable request.

\begin{appendix}

\section{The Lagrangian and the Hamiltonian  of the JTL}
\label{lagham}

Let us rederive (\ref{ave7})  in the framework of  Lagrange approach.
The Lagrangian ${\cal L}$ is
\begin{eqnarray}
\label{lag}
&&{\cal L}=\frac{\ell}{2}\sum_n\left(\frac{dq_n}{dt}\right)^2
+\frac{c_2\hbar^2}{8e^2}\sum_n\left(\frac{d\varphi_n}{dt}\right)^2\\
&&-\frac{1}{2c}\sum_n \left(q_{n}-q_{n+1}\right)^2
+\frac{\hbar}{2e}I_c\sum_n\cos\varphi_n+\frac{\hbar}{2e}\sum_n
\frac{dq_n}{dt}\varphi_n \,.\nonumber
\end{eqnarray}
Lagrange equations have the form
\begin{eqnarray}
\frac{d}{dt}\left(\frac{\partial {\cal L}}{\partial \dot{Q}}\right)-\frac{\partial {\cal L}}{\partial Q}=0,
\end{eqnarray}
where $Q$ is any of the dynamical variables.
Lagrange equation corresponding to $q_n$
\begin{eqnarray}
\label{canonic8}
\ell\frac{d^2q_n}{dt^2}+\frac{\hbar}{2e}\frac{d}{dt}\varphi_n
+ \frac{1}{c}\left(2q_{n}-q_{n+1}-q_{n-1}\right)=0\,,
\end{eqnarray}
reproduces (\ref{ave7a}).
Lagrange equation corresponding to $\varphi_n$
\begin{eqnarray}
\label{canon}
\frac{\hbar^2c_2}{4e^2}\sum_n\frac{d^2\varphi_n}{dt^2}+\frac{\hbar}{2e}I_c\sin\varphi_n
-\frac{\hbar}{2e}\frac{dq_n}{dt}=0\,,
\end{eqnarray}
reproduces  (\ref{ave7b}).

In the continuum limit the Lagrangian (\ref{lag}) is
\begin{eqnarray}
\label{lag2}
{\cal L}=\int dx\left[\frac{L}{2}\left(\frac{\partial q}{\partial t}\right)^2
 +\frac{C_2E_J^2}{2}\left(\frac{\partial \varphi}{\partial t}\right)^2\right.\nonumber\\
\left.-\frac{1}{2C}\left(\frac{\partial q}{\partial x}\right)^2+E_JI_c\cos\varphi
+E_J\frac{\partial q}{\partial t}\varphi \right]\,.
\end{eqnarray}
The Hamiltonian, corresponding to the Lagrangian (\ref{lag2}), contains two pairs of conjugate variables $(\pi,q)$ and $(p,\varphi)$ and has the form
\begin{eqnarray}
\label{ham4}
{\cal H}=\int dx\left[\frac{\left(\pi-E_J\varphi\right)^2}{2L}+\frac{p^2}{2C_2E_J^2}
\right.\nonumber\\
\left.+\frac{1}{2C}\left(\frac{\partial q}{\partial x}\right)^2-E_JI_c\cos\varphi\right]\,.
\end{eqnarray}
Hamilton equations have the form
\begin{subequations}
\label{he}
\begin{alignat}{4}
\frac{\partial Q}{\partial t}&=\frac{\partial {\cal H}}{\partial P} \, ,\label{hea}\\
\frac{\partial P}{\partial t}&=\frac{\partial}{\partial x}\frac{\partial {\cal H}}{\partial(\partial Q/\partial x)} -\frac{\partial {\cal H}}{\partial Q}\, ,\label{heb}
\end{alignat}
\end{subequations}
where $(P,Q)$ is any pair of the conjugate variables. Hamilton equations corresponding to $(\pi,q)$
\begin{subequations}
\label{he2}
\begin{alignat}{4}
\frac{\partial q}{\partial t}&=\frac{\pi-E_J\varphi}{L}\, ,\label{he2a}\\
\frac{\partial \pi}{\partial t}&=\frac{1}{C}\frac{\partial^2 q}{\partial x^2}\, ,\label{he2b}
\end{alignat}
\end{subequations}
reproduce  (\ref{wave7xa}).
Hamilton equations corresponding to $(p,\varphi)$
\begin{subequations}
\label{he3}
\begin{alignat}{4}
\frac{\partial \varphi}{\partial t}&=\frac{p}{C_2E_J^2} \, ,\label{he3a}\\
\frac{\partial p}{\partial t}&
=\frac{E_J}{L}\left(\pi-E_J\varphi\right)-E_JI_c\sin\varphi  \, ,\label{he3b}
\end{alignat}
\end{subequations}
reproduce  (\ref{wave7xb}), if we take into account (\ref{he2a}).

For $c_2=0$, taking into account  (\ref{canon}), we can write down the Lagrangian
(\ref{lag}) in the form which doesn't contain $\varphi_n$
\begin{eqnarray}
\label{lagg}
{\cal L}=\frac{\ell}{2}\sum_n\left(\frac{dq_n}{dt}\right)^2
-\frac{1}{2c}\sum_n \left(q_{n}-q_{n+1}\right)^2\nonumber\\
+\frac{\hbar}{2e}I_c\sum_n\sqrt{1-\left(\frac{1}{I_c}\frac{dq_n}{dt}\right)^2}\nonumber\\
+\frac{\hbar}{2e}\sum_n
\frac{dq_n}{dt}\cdot\sin^{-1}\left(\frac{1}{I_c}\frac{dq_n}{dt}\right) \,.
\end{eqnarray}
One can easily check up that (\ref{ave9}) is the Lagrange equation corresponding to the Lagrangian (\ref{lagg}).

If we assume additionally that $\ell=0$, the  Hamiltonian, corresponding to the Lagrangian (\ref{lagg}), has a neat form
\begin{eqnarray}
\label{hamm}
{\cal H}=\frac{1}{2c}\sum_n \left(q_{n}-q_{n+1}\right)^2-
\frac{\hbar}{2e}I_c\sum_n\cos\left(\frac{2ep_n}{\hbar}\right) \,.\nonumber\\
\end{eqnarray}

\section{Travelling wave in the discrete JTL}
\label{spread}

In the particular case $c_2=0$ we can eliminate $q_n$  from (\ref{ave7}) and obtain closed equation for $\varphi_n$ in the form
\begin{eqnarray}
\label{eq0}
\frac{I_c}{ c}\left(\sin\varphi_{n+1}+\sin\varphi_{n-1}-2\sin\varphi_n\right)\nonumber\\
=I_c\ell\frac{d^2\sin\varphi_n}{dt^2}+\frac{\hbar }{2e}
\frac{d^2\varphi_n }{dt^2}\,.
\end{eqnarray}
Travelling wave solution of  (\ref{eq0}) has the form
\begin{eqnarray}
\label{cano}
\varphi_n(t)=\varphi(t-n\tau)\,,
\end{eqnarray}
where  $\varphi(t)$ is some unknown functions, and $\tau$ is the parameter determining the velocity of the travelling wave. For such solution,  (\ref{eq0}) takes the form
\begin{eqnarray}
\label{eq}
\frac{I_c}{ c}\left[\sin\varphi(t+\tau) +\sin\varphi(t-\tau)-2\sin\varphi(t)\right]\nonumber\\ =I_c\ell\frac{d^2\sin\varphi(t)}{dt^2}+\frac{\hbar }{2e}
\frac{d^2\varphi(t) }{dt^2}\,.
\end{eqnarray}
We are interested in the solution of  (\ref{eq})  satisfying boundary conditions
\begin{eqnarray}
\lim_{t\to-\infty}\varphi(t)=\varphi_2 \,,\hskip .5cm \lim_{t\to+\infty}\varphi(t)=\varphi_1\,.
\end{eqnarray}
Note, that if we twice integrate (\ref{eq}) with respect to $t$, we obtain equation
\begin{eqnarray}
\label{eq2}
\frac{I_c}{ c}\tau^2\Delta\sin\varphi=I_c\ell\Delta\sin\varphi+\frac{\hbar }{2e}
\Delta\varphi\,,
\end{eqnarray}
which, taking into account that $U=\Lambda/\tau$,  reproduces  (\ref{velocity}).

\section{The method of time averaging: weak shocks}
\label{integral}

\begin{figure}[h]
\includegraphics[width=\columnwidth]{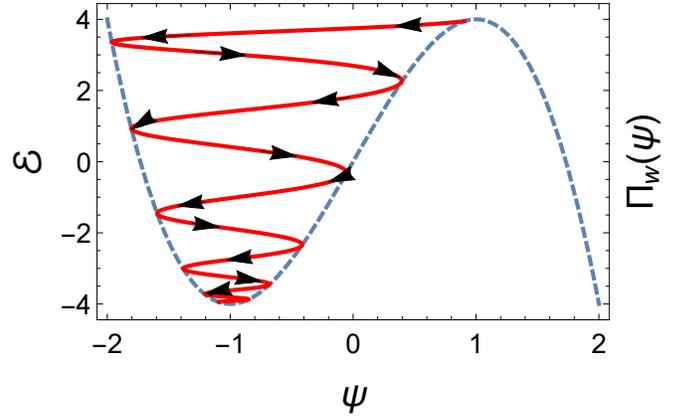}
\caption{The particle trajectory in phase space  according to Eqs. (\ref{psih}) and (\ref{sim}) (solid red curve).}
\label{weab}
\end{figure}

We want to show how the method of time averaging works on the simplest possible example,
applying it to  (\ref{dvec}) and
considering, in addition to weak damping, the case of weak
shock. We
make the change of  variable
\begin{eqnarray}
\label{tr}
\psi=\frac{2(\varphi-\varphi_0)}{\Delta\varphi},
\end{eqnarray}
where $\varphi_0=(\varphi_1+\varphi_2)/2$, so that
$t=-\infty$ state would correspond to $\psi=1$, and   $t=+\infty$ state - to $\psi=-1$.
After we  rescale   in comparison to (\ref{change})
$\Phi\tau\to\tau$, $(\gamma_2+\gamma_1\cos\varphi_0)/\Phi\to\gamma$
$(\Phi=\sqrt{u_0^2\tan\varphi_0\Delta\varphi/24})$,
 (\ref{dvec}) takes the form
\begin{eqnarray}
\label{dvec3b}
\frac{d^2 \psi}{d\tau^2}+\gamma\frac{d \psi}{d\tau}+6(1-\psi^2)=0\,.
\end{eqnarray}
Equation (\ref{eee}) in our case becomes
\begin{eqnarray}
\label{eeey5}
{\cal A}({\cal E})=2\gamma\frac{\int d\varphi \sqrt{{\cal E}-\Pi(\varphi)}}
{\int d\varphi/\sqrt{{\cal E}-\Pi(\varphi)}}=2\gamma <{\cal E}_{kin}>\,.
\end{eqnarray}
The potential energy
\begin{eqnarray}
\label{we}
\Pi_w(\psi)=-2\psi^3+6\psi
\end{eqnarray}
is presented on Fig. \ref{weab}.
It    has local minimum $\Pi^{(min)}_{w}=-4$ at $\psi=-1$ and local maximum $\Pi^{(max)}_{w}=4$ at $\psi=1$.
Equation (\ref{con}) in the present case,
\begin{eqnarray}
\label{w34}
\left(\frac{d\psi_{ud}}{d\tau}\right)^2=4\psi_{ud}^3-12\psi_{ud}+2{\cal E}\,,
\end{eqnarray}
defines Weierstrass elliptic function (with $g_2=12$). \cite{abram}
\begin{eqnarray}
\label{gy}
\psi_{ud}(\tau;{\cal E})={\cal P}(\tau;12,-2{\cal E})\,.
\end{eqnarray}
Thus the damped solution is
\begin{eqnarray}
\label{psih}
\psi(\tau)={\cal P}(\tau;12,-2{\cal E}(\tau))\,.
\end{eqnarray}

Let us make a short cut in the  method of time averaging, by assuming (being inspired by the example of harmonic oscillator)
\begin{eqnarray}
\label{eee27}
<{\cal E}_{kin}>={\cal E}-\Pi_w^{(min)}\,.
\end{eqnarray}
After that,  (\ref{eee}) is easily solved
\begin{eqnarray}
\label{sim}
{\cal E}=\Pi^{(min)}_{w}+\left(\Pi^{(max)}_{w}
-\Pi^{(min)}_{w}\right)e^{-2\gamma\tau}\,.
\end{eqnarray}

Now let us calculate ${\cal E}_{kin}$ in earnest.
The integrals entering into (\ref{eeey5}) are elliptic:
\begin{subequations}
\label{ellip}
\begin{alignat}{4}
&Y_0({\cal E})=\int_c^b \frac{d\psi}{\sqrt{P(\psi)}}\,,\;\;\;
{\cal N}({\cal E})=\int_c^b d\psi \sqrt{P(\psi)},\label{ellipa}\\
&P(\psi)=2\psi^3-6\psi+{\cal E} \, ,\label{ellipb}
\end{alignat}
\end{subequations}
where $a,b,c$ ($a>b>c$)
 are  the  roots of  cubic  equation $P(\psi)=0$.
The first integral in  (\ref{ellipa}) is a table integral \cite{grad}.
 The second integral has to be calculated.

For the theory of elliptic integrals one may turn to excellent book by E. Goursat  \cite{goursat}
The book not only formulates the theorem which will be important for us:

\noindent
All integrals
\begin{eqnarray}
Y_m=\int\frac{\psi^m d\psi}{\sqrt{P(\psi)}}\,,
\end{eqnarray}
where $m$ is an arbitrary natural number and $P(\psi)$ is some polynomial of power $p$,
are expressed through the  $p-1$ first integrals $Y_0,Y_1,\dots,Y_{p-2}$ and algebraic quantities,

\noindent
but shows how the reduction should be made in practice.

So  let's turn to calculation of ${\cal N}$.
Integrating the identity
\begin{eqnarray}
\sqrt{P(\psi)}=\frac{{\cal E}-6\psi+2\psi^3}{\sqrt{P(\psi)}}
\end{eqnarray}
we obtain
\begin{eqnarray}
\label{gour}
{\cal N}={\cal E}Y_0-6Y_1+2Y_3\,.
\end{eqnarray}
$Y_0$ and $Y_1$ are table integrals \cite{grad}:
\begin{subequations}
\label{gou}
\begin{alignat}{4}
Y_0&=\frac{2}{\sqrt{a-c}}K(k) \, ,\label{gou1}\\
Y_1&=\frac{2a}{\sqrt{a-c}}K(k)-2\sqrt{a-c}E(k) \,,\label{gou2}
\end{alignat}
\end{subequations}
where $K$ and $E$ are complete elliptic integrals of the first  and second kind respectively, and $k=\sqrt{(b-c)/(a-c)}$.
$Y_2$  can be expressed through $Y_0$ and $Y_1$ \cite{goursat}.
Integrating the identity
\begin{eqnarray}
\frac{d}{d\psi}\left[\psi\sqrt{P(\psi)}\right]=\sqrt{P(\psi)}
-\frac{3(\psi-\psi^3)}{\sqrt{P(\psi)}},
\end{eqnarray}
we obtain
\begin{eqnarray}
\label{gourd}
{\cal N}-3Y_1+3Y_3=0\,.
\end{eqnarray}
Combining (\ref{gour})   and (\ref{gourd}) we obtain
\begin{eqnarray}
\label{gg}
{\cal N}&=&\frac{3}{5}\left({\cal E}Y_0-4Y_1\right)\\
&=&\frac{6}{5\sqrt{a-c}}\left[\left({\cal E}-4a\right)K(k)+4(a-c)E(k)\right]\,,\nonumber
\end{eqnarray}
and, finally,
\begin{eqnarray}
\label{ou}
<{\cal E}_{kin}>=\frac{3}{5}\left[{\cal E}-4a+4(a-c)\frac{E(k)}{K(k)}\right]\,.
\end{eqnarray}
(One should keep in mind that $a,c,k$ are functions of ${\cal E}$.)
\begin{figure}[h]
\includegraphics[width=\columnwidth]{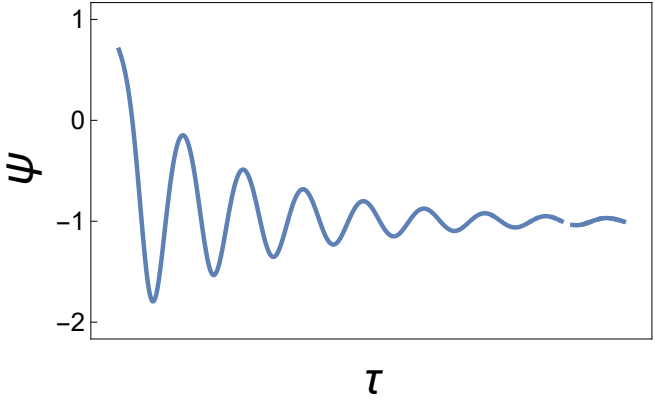}
\caption{Coordinate dependence of the fictitious particle coordinate according to Eqs. (\ref{ou}), (\ref{eeey5}), (\ref{ee}), and (\ref{psih}).}
\label{wec}
\end{figure}

Let us analyse the limiting cases of (\ref{ou}).
Obviously, averaged kinetic energy should go to zero both when ${\cal E}\to 4$,
 because the period of oscillations  goes to infinity, and when ${\cal E}\to -4$, because the particle approaches the bottom of the well. Equation (\ref{ou}) clearly demonstrates such behavior. To check it up it is enough to inspect Fig. \ref{weab} and keep in mind that
\begin{subequations}
\begin{alignat}{4}
\lim_{k\to 0}K(k)& =\lim_{k\to 0}E(k)=\frac{\pi}{2} \, ,\\
\lim_{k\to 1}K(k)&=\infty\,,\;\;\;\; \lim_{k\to 1}E(k)=1\, .
\end{alignat}
\end{subequations}

Averaged kinetic energy being found, we can easily calculate integral (\ref{ee}) numerically. The solution obtained in the result of the approximation is presented on Fig. \ref{wec}.
The shock front was not presented  on purpose.
The method of averaging is meaningful, provided the time scale of the change of energy is much larger than the dynamical time scale (inverse frequency of oscillations).
When the energy is close to $\Pi_w^{(max)}$,
the period of oscillation is very large, and the method ceases to be applicable.

\begin{figure}[h]
\includegraphics[width=\columnwidth]{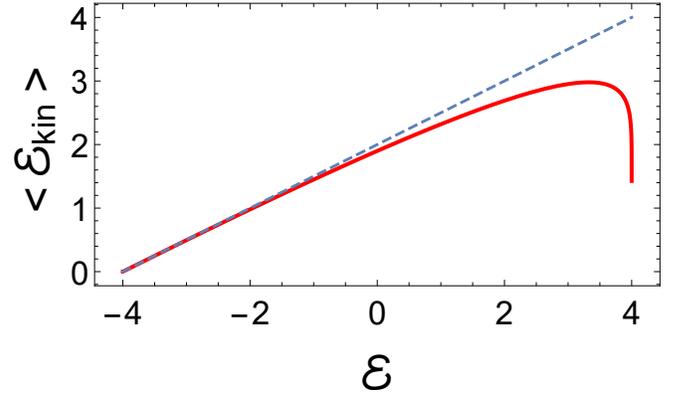}
\caption{The averaged kinetic energy of the fictitious particle according to  (\ref{ou})  (solid red curve) and  according to  (\ref{eee27}) (dashed blue curve).}
\label{bottom}
\end{figure}

Actually, when the method of averaging is applicable, the complicated
result (\ref{ou}) coincides with the  naive approximation (\ref{eee27}). To show it, we plot
both equations on Fig. \ref{bottom}. The results are close
 everywhere, apart from the  vicinity of $\Pi_w^{(max)}$.
So (\ref{psih}) and (\ref{sim}) give good and simple analytic approximation
to the profile of the shock wave valid everywhere, apart from the vicinity of the shock front.

\section{The symmetry of the modified Burgers equation}
\label{exact}

By trivial change of variables we can transform (\ref{proba22}) to
\begin{eqnarray}
\label{proba2}
u_t + uu_x+ uu_{tx}=0\,.
\end{eqnarray}
To warm up, let us copy from Refs. \cite{olver,arrigo} the symmetry analysis of Burgers equation
\begin{eqnarray}
\label{2.57}
u_{t}+uu_x=u_{xx}.
\end{eqnarray}
 The symmetry group of (\ref{2.57}) is generated by the vector field
\begin{eqnarray}
{\bf v}=T(t,x,u)\frac{\partial}{\partial t}+X(t,x,u)\frac{\partial}{\partial x}+U(t,x,u)\frac{\partial}{\partial u}
\end{eqnarray}
and its first and the second prolongations
\begin{subequations}
\label{prolong}
\begin{alignat}{4}
\text{pr}^{(1)}{\bf v}&={\bf v}+U^t\frac{\partial}{\partial u_t}+U^x\frac{\partial}{\partial u_x} \,,\label{prolonga}\\
\text{pr}^{(2)}{\bf v}&=\text{pr}^{(1)}{\bf v}
+U^{tt}\frac{\partial}{\partial u_{tt}}+U^{tx}\frac{\partial}{\partial u_{tx}}
+U^{xx}\frac{\partial}{\partial u_{xx}}\,. \label{prolongb}
\end{alignat}
\end{subequations}
To write down (\ref{prolong}) explicitly,
we define total derivatives
\begin{subequations}
\label{der}
\begin{alignat}{4}
D_t&=\frac{\partial}{\partial t}+u_t\frac{\partial}{\partial u}+
u_{tt}\frac{\partial}{\partial u_t}+u_{tx}\frac{\partial}{\partial u_x}+\dots\,,\\
D_x&=\frac{\partial}{\partial x}+u_x\frac{\partial}{\partial u}+
u_{tx}\frac{\partial}{\partial u_t}+u_{xx}\frac{\partial}{\partial u_x}+\dots\,.
\end{alignat}
\end{subequations}
The coefficients of the first prolongation are
\begin{subequations}
\label{2.45}
\begin{alignat}{4}
U^t&=D_tU-u_tD_tT-u_xD_tX \\
U^x&=D_xU-u_tD_xT-u_xD_xX\,.
\end{alignat}
\end{subequations}
The coefficients entering the second prolongation are
\begin{subequations}
\label{2.4}
\begin{alignat}{4}
U^{tt}&=D_tU^t-u_{tt}D_tT-u_{tx}D_tX\,,\\
U^{tx}&=D_tU^x-u_{tx}D_tT-u_{xx}D_tX\,,\\
      &=D_xU^t-u_{tt}D_xT-u_{tx}D_xX\,,\\
U^{xx}&=D_xU^x-u_{tx}D_xT-u_{xx}D_xX\,.
\end{alignat}
\end{subequations}
Differentiating we obtain
\begin{subequations}
\begin{alignat}{4}
U^t&=U_t+u_tU_u-u_t(T_t+u_tT_u)-u_x(X_t+u_tX_u) \\
U^x&=U_x+u_xU_u-u_t(T_x+u_xT_u)-u_x(X_x+u_xX_u)\,,
\end{alignat}
\end{subequations}
and
\begin{eqnarray}
\label{2.46}
&&U^{xx}=U_{xx}-u_tT_{xx}+u_x\left(2U_{xu}-X_{xx}\right)\nonumber\\
&&-2u_tu_xT_{xu}+u_x^2\left(U_{uu}-2X_{xu}\right)-u_tu_x^2T_{uu}-u_x^3X_{uu}\nonumber\\
&&-2u_{tx}T_x+u_{xx}\left(U_{u}-2X_{x}\right)\nonumber\\
&&-2u_xu_{tx}T_u-u_tu_{xx}T_u-3u_xu_{xx}X_u.
\end{eqnarray}

Applying the second prolongation pr$^2{\bf v}$ to (\ref{2.57}) we find that $T,X,U$ must satisfy
the symmetry condition
\begin{eqnarray}
U^t+uU^{x}+u_xU=U^{xx}.
\end{eqnarray}
Substituting (\ref{2.45}) and (\ref{2.46}) eventually
leads to the following set of determining equations:
\begin{eqnarray}
\label{3.51}
T_x=0,\;\;\;T_u=0,\;\;\;X_u=0,\;\;\;U_{uu}=0,\nonumber\\
2X_x-T_t=0,\;\;\;U_t+uU_x-U_{xx}=0,\\
X_t-X_{xx}-uX_x-U+2U_{xu}=0.\nonumber
\end{eqnarray}
Solving the first 5 equations of (\ref{3.51}) gives
\begin{subequations}
\label{3.52}
\begin{alignat}{4}
X&=\frac{1}{2}T'(t)x+A(t) \, , \\
U&=B(t,x)u+C(t,x)\,,
\end{alignat}
\end{subequations}
where $A(t)$, $B(t,x)$ and $C(t,x)$ are arbitrary functions. Substituting
(\ref{3.52}) into the remaining two equations of (\ref{3.51}), isolating coefficients
with respect to u and solving gives the infinitesimals:
\begin{subequations}
\begin{alignat}{4}
T&=c_0+2c_1t+c_2t^2 \, , \\
X&=c_3+c_4t+c_1x+c_2tx \, , \\
U&=-(c_2t+c_1)u+c_2x+c_4\,.
\end{alignat}
\end{subequations}

Now let us come to the symmetry analysis of (\ref{proba2}).
Here we need
\begin{eqnarray}
\label{2.47}
&&U^{tx}=U_{tx}+u_t(U_{xu}-T_{tx})+u_x(U_{tu}-X_{tx})\nonumber\\
&&+u_tu_x(U_{uu}-T_{xu}-X_{xu})-u_t^2T_{xu}-u_x^2X_{tu}\nonumber\\
&&-u_t^2u_xT_{uu}-u_tu_x^2X_{uu}\\
&&-u_{tt}T_x+u_{tx}(U_u-T_t-X_x)-u_{xx}X_t\nonumber\\
&&-u_{tt}u_xT_u-2u_{tx}(u_tT_u+u_xX_u)-u_tu_{xx}X_u\,.\nonumber
\end{eqnarray}
Applying the second prolongation pr$^2{\bf v}$ to (\ref{proba2}) we find that $T,X,U$ must satisfy
the symmetry condition
\begin{eqnarray}
U^t+uU^{x}+u_xU+u_{tx}U+uU^{tx}=0.
\end{eqnarray}
Substituting (\ref{2.45}) and (\ref{2.47}) eventually
leads to the following set of determining equations:
\begin{eqnarray}
\label{3.51b}
T_u=0,\;\;\;X_u=0,\;\;\;X_t=0,\;\;\;T_x&=&0,\;\;\;U_{uu}=0,\nonumber\\
U_{tu}+T_t=0,\;\;\;U_t+uU_{tx}+uU_x&=&0,\\
U-u^2U_{xu}-uX_x&=&0\;.\nonumber
\end{eqnarray}
Solving the first 5 equations of (\ref{3.51b}) gives
\begin{subequations}
\label{3.52b}
\begin{alignat}{4}
U&=-T(t)u+A(x)u+B(x,t) \, , \\
X&=X(x)\,.
\end{alignat}
\end{subequations}
Substituting
(\ref{3.52b}) into the remaining two equations of (\ref{3.51b})
we obtain
\begin{subequations}
\begin{alignat}{4}
-T'u+B_t+uB_{tx}+u^2A'&=0\,,\\
-T(t)u+A(x)u+B(x,t)-u^2A'-uX'&=0\,.
\end{alignat}
\end{subequations}
Isolating coefficients
with respect to u and solving gives the infinitesimals:
\begin{subequations}
\begin{alignat}{4}
T&=c_0 \, , \\
X&=c_1+c_2x \, , \\
U&=c_2u\,.
\end{alignat}
\end{subequations}
The symmetry of the mBE turned out to be rather low.
Apart from time and space translations, the only symmetry of the equation is with respect to transformation
\begin{eqnarray}
u\rightarrow Cu,\;\;x\rightarrow Cx,\;\;t\rightarrow t\,.
\end{eqnarray}
This symmetry was obvious by inspection of the mBE. What was presented above, is the proof that the equation does not have any other classical symmetries.
It would be interesting to check up mBE for the nonclassical symmetries.

Taking into account the symmetry of the modified Burgers equation, we may look for an exact solution of  (\ref{proba2}) in the form
\begin{eqnarray}
\label{ex}
u(x,t)=xv(t)\,.
\end{eqnarray}
Substituting (\ref{ex}) into (\ref{proba2}) we obtain ordinary differential equation for $v(t)$
\begin{eqnarray}
\frac{1}{v}\frac{dv}{dt}+v+\nu\frac{dv}{dt}=0\,,
\end{eqnarray}
with the solution
\begin{eqnarray}
t=\frac{1}{v}-\nu\ln|v|\,.
\end{eqnarray}

\section{Discontinuities in the JTL with ohmic resistors in series with the JJ}
\label{ser}

Consider again the simple wave approximation  for the dissipationless JTL (\ref{ve8}).
The characteristic equations  for  the right going wave are \cite{whitham}
\begin{subequations}
\label{char}
\begin{alignat}{4}
\frac{dI}{dt} &= 0 \, ,\label{chara}\\
\frac{dx}{dt}&=u(I) \, .\label{charb}
\end{alignat}
\end{subequations}
The solution of  (\ref{char}) for the Cauchy problem is
\begin{subequations}
\label{char2}
\begin{alignat}{4}
I(x,t)&=I(\xi,0) \, ,\label{char2a}\\
x&=\xi+u(I(\xi,0))t \, ,\label{char2b}
\end{alignat}
\end{subequations}
where $I(\xi,0)$ is given by the initial condition for the problem.

Consider now the JTL with ohmic dissipation, described by  (\ref{ve9d}).
The characteristic equations  for the right going wave are \cite{whitham}
\begin{subequations}
\label{par}
\begin{alignat}{4}
\frac{dI}{dt} &= -\mu u^2(I)I \, ,\label{para}\\
\frac{dx}{dt}&=u(I) \, .\label{parb}
\end{alignat}
\end{subequations}
Equation (\ref{par})  shows that the
initial value of $I(\xi,0)$ at a given point $\xi$ is propagating  along the characteristic given by equation
\begin{eqnarray}
\label{char3b}
x=\xi+\int_0^tu(I(\xi,t'))dt'\,,
\end{eqnarray}
decreasing in the process of propagation according to  (\ref{para}).
In  (\ref{char3b}), $I(\xi,t)$ is the solution of  (\ref{para})
with the same  initial condition as before.

Equation (\ref{char3b}) (and its particular case (\ref{char2b})) allow us to understand why the discontinuities in $I(x,t)$ are formed in the solution
starting with the continuous initial condition $I(x,0)$. Values of current, corresponding to larger  $I^2$  propagate slower than those corresponding to smaller $I^2$
(note that $u(I)/dI^2<0$ for all $I$), so if there are $x$ intervals, where
$I^2(x,t)$ increases with $x$, the characteristics diverge.
The  discontinuities in $I(x,t)$  correspond to the crossing of the characteristic curves, that is to the existence of their envelope \cite{whitham},
 satisfying simultaneously  (\ref{char3b}) and equation
\begin{eqnarray}
\label{bre}
0=1+\int_0^t\left.\frac{d u}{d I^2}\right|_{I= I(\xi,t')}\frac{\partial I^2(\xi,t')}{\partial \xi}dt'\,.
\end{eqnarray}
The discontinuities appear at minimal $t$ for which  (\ref{bre}) has a solution .

 In the absence of ohmic dissipation,
 $I(\xi,t)$ is time independent, so if there are $\xi$ intervals where
$I^2(\xi,0)$ increases with $\xi$,  (\ref{bre}) has a solution for $t$ large enough, however small   $d I^2(\xi,0)/d \xi >0$ is. In the presence of dissipation,  the current decreases exponentially with time, so  (\ref{bre}) has a solution only for steep enough current rises in the initial condition.
Geometrically, because in the presence of dissipation the system becomes more and more linear with time,  the characteristics become more and more parallel, and don't necessarily have to cross, in distinction to the case of no dissipation.

\end{appendix}

\end{document}